\documentclass{aastex63}

\usepackage{subfig}
\usepackage{graphicx}
\usepackage{float}
\usepackage{bm} 
\usepackage{amsmath} 

\begin{document}

\submitjournal{ApJ}

\shorttitle{Correcting Transiting Exoplanet Light Curves for Stellar Spots}
\shortauthors{N. Nikolaou et al.}

\title{Lessons Learned from the 1st ARIEL Machine Learning Challenge:\\Correcting Transiting Exoplanet Light Curves for Stellar Spots}

\correspondingauthor{Nikolaos Nikolaou}
\email{n.nikolaou@ucl.ac.uk}

\author[0000-0001-8453-7574]{Nikolaos Nikolaou}
\affiliation{Department of Physics and Astronomy, University College London, Gower Street, London, WC1E 6BT, UK}
\author[0000-0002-4205-5267]{Ingo P. Waldmann}
\affiliation{Department of Physics and Astronomy, University College London, Gower Street, London, WC1E 6BT, UK}
\author[0000-0003-3840-1793]{Angelos Tsiaras}
\affiliation{Department of Physics and Astronomy, University College London, Gower Street, London, WC1E 6BT, UK}
\author[0000-0001-8587-2112]{Mario Morvan}
\affiliation{Department of Physics and Astronomy, University College London, Gower Street, London, WC1E 6BT, UK}
\author[0000-0002-5494-3237]{Billy Edwards}
\affiliation{Department of Physics and Astronomy, University College London, Gower Street, London, WC1E 6BT, UK}
\author[0000-0002-9616-1524]{Kai Hou Yip}
\affiliation{Department of Physics and Astronomy, University College London, Gower Street, London, WC1E 6BT, UK}
\author[0000-0001-6058-6654]{Giovanna Tinetti}
\affiliation{Department of Physics and Astronomy, University College London, Gower Street, London, WC1E 6BT, UK}
\author{Subhajit Sarkar}
\affiliation{School of Physics and Astronomy, Cardiff University, The Parade, Cardiff, CF24 3AA, UK}
\author{James M. Dawson}
\affiliation{School of Physics and Astronomy, Cardiff University, The Parade, Cardiff, CF24 3AA, UK}
\author{Vadim Borisov}
\affiliation{Department of Computer Science, University of Tuebingen, Tuebingen, Germany}
\author{Gjergji Kasneci}
\affiliation{Department of Computer Science, University of Tuebingen, Tuebingen, Germany}
\author{Matej Petkovi\'{c}}
\affiliation{Jo\v{z}ef Stefan Institute, Ljubljana, Slovenia}
\author{Toma\v{z} Stepi\v{s}nik}
\affiliation{Jo\v{z}ef Stefan Institute, Ljubljana, Slovenia}
\author{Tarek Al-Ubaidi}
\affiliation{DCCS, Austria}
\affiliation{Space Research Institute, Austrian Academy of Sciences, Austria}
\author[0000-0003-2021-6557]{Rachel Louise Bailey}
\affiliation{Space Research Institute, Austrian Academy of Sciences, Austria}
\author[0000-0003-3566-5507]{Michael Granitzer}
\affiliation{Chair of Data Science, University of Passau, Germany}
\author[0000-0002-2952-5519]{Sahib Julka}
\affiliation{Chair of Data Science, University of Passau, Germany}
\author{Roman Kern}
\affiliation{Know-Center GmbH - Research Center for Data-Driven Business \& Big Data Analytics, Austria}
\author[0000-0001-7169-4300]{Patrick Ofner}
\affiliation{Know-Center GmbH - Research Center for Data-Driven Business \& Big Data Analytics, Austria}
\author{Stefan Wagner}
\affiliation{Commission for Astronomy, Austrian Academy of Sciences, Graz, Austria}
\author[0000-0002-7726-7868]{Lukas Heppe}
\affiliation{Artificial Intelligence Group, TU Dortmund University, Germany}
\author[0000-0002-5515-6278]{Mirko Bunse}
\affiliation{Artificial Intelligence Group, TU Dortmund University, Germany}
\author[0000-0003-1153-5986]{Katharina Morik}
\affiliation{Artificial Intelligence Group, TU Dortmund University, Germany}











\begin{abstract}

The last decade has witnessed a rapid growth of the field of exoplanet discovery and characterisation. However, several big challenges remain, many of which could be addressed using machine learning methodology. For instance, the most prolific method for detecting exoplanets and inferring several of their characteristics, transit photometry, is very sensitive to the presence of stellar spots. The current practice in the literature is to identify the effects of spots visually and correct for them manually or discard the affected data. This paper explores a first step towards fully automating the efficient and precise derivation of transit depths from transit light curves in the presence of stellar spots. The methods and results we present were obtained in the context of the 1st Machine Learning Challenge organized for the European Space Agency’s upcoming Ariel mission. We first present the problem, the simulated Ariel-like data and outline the Challenge while identifying best practices for organizing similar challenges in the future. Finally, we present the solutions obtained by the top-5 winning teams, provide their code and discuss their implications. Successful solutions either construct highly non-linear (w.r.t. the raw data) models with minimal preprocessing --deep neural networks and ensemble methods-- or amount to obtaining meaningful statistics from the light curves, constructing linear models on which yields comparably good predictive performance. 

\end{abstract}

\keywords{ARIEL, exoplanets, transit photometry, light curves, stellar spots, machine learning}


\section{Introduction} \label{sec:intro}

In the coming decade, exoplanet atmospheric spectroscopy will undergo a revolution with a number of upcoming space and ground-based instruments providing unprecedented amounts of high-quality data. Most notable are of course the Extremely Large Telescopes \citep[e.g.][]{gilmozzi_european_2007, nelson_status_2008,johns_giant_2012} on the ground and the James Webb Space Telescope \citep{gardner_james_2006} and the Ariel space telescope \citep{tinetti_science_2016}. One of the outstanding challenges to high-precision spectrophotometry of exoplanets is the presence of stellar noise. Here we will address in particular the presence of occulted star spots in the spectro-photometric light curves of the Ariel space mission. 
The chromatic dependence of spots and faculae can adversely affect the measured exoplanetary transmission spectrum (through a biasing of the derived transit depth) as well as affect other lightcurve paramters, such as limb-darkening, mid-transit times. This is discussed in detail in 
\citep[e.g.][and references therein]{sing_continuum_2015,nikolov_hubble_2013,rabus_cool_2009,mccullough_water_2014,rackham_transit_2018,rackham_transit_2019,zellem_forecasting_2017,iyer_influence_2020}. There exists a large body of literature on modelling star spot signatures in photometric and radial velocity data 
\citep[e.g.][]{boisse_soap_2012,dumusque_planetary_2011,dumusque_soap_2014,lanza_deriving_2011,aigrain_simple_2012,herrero_modelling_2016,zhao_fiesta_2020,gilbertson_towards_2020,lisogorskyi_impact_2020}. Recently, \citet{rosich_correcting_2020} proposed a correction of the chromatic effects using Bayesian inverse modelling of long duration spectro-photometric time-series data with promising results. 

In this publication, we explore the use of machine learning techniques to detect and correct for spot crossings in simulated data of the Ariel space mission. In particular, we report on the top five results of the 1st Ariel Mission Machine Learning Challenge (henceforth: the Challenge), which was concerned with the task of \emph{correcting transiting exoplanet light curves for the presence of stellar spots}. The primary goal of the Challenge was thus to investigate if machine learning approaches are in principle suited to correcting star spot crossings in spectro-photometric lightcurves across a large range of stellar and planetary parameters as well as observational signal to noise regimes. 

To date, the use of machine learning approaches in exoplanets is still nascent but a burgeoning interest has seen the successful application of machine learning --and deep learning, in particular-- to a variety of exoplanetary problems. These include (but are not limited to) the detection of exoplanet transits in survey data \citep[e.g.][]{shallue_identifying_2018,pearson_searching_2018,osborn_rapid_2020}, the predictive modelling of planetary parameters 
\citep{lam_kipping18,alibert_using_2019}, instrument de-trending 
\citep[e.g.][]{waldmann_cocktail,morello_new_2014,gibson_gaussian_2012,morvan_detrending_2020} and the modelling and retrieval of atmospheric spectra 
\citep[e.g.][]{waldmann_dreaming_2016,marquez-neila_supervised_2018,zingales_exogan_2018, cobb_ensemble_2019, nixon_assessment_2020, himes_accurate_2020}. 

As many problems in the field of exoplanetary science, the issue of star spot crossings is characterised by a combination of challenges: (i) a large amount of data to process\footnote{As Ariel is an upcoming space mission, the data in our case are obtained via simulations.}, (ii) low signal to noise ratio, (iii) an underlying pattern which is non-linear and whose parametric form is a-priori unknown, (iv) the available information comes in multiple forms (time dependent and independent), and finally (v) a high degree of degeneracy. These issues are commonly addressed by machine learning approaches.

This takes us to the second objective of the Challenge: promoting the interaction between the astrophysics and the machine learning communities. To this end, the Challenge targeted both audiences by being officially organized in the context of the ECML-PKDD 2019 conference\footnote{ECML-PKDD, the European Conference on Machine Learning and Principles and Practice of Knowledge Discovery in Databases, is one of the leading academic conferences on machine learning and knowledge discovery, held in Europe every year.} and also having a strong presence in the joint EPSC-DPS 2019\footnote{The European Planetary Science Congress (EPSC) and the American Astronomical Society's Division of Planetary Science (DPS) held a Joint Meeting at 2019.} conference via a dedicated session. The Challenge ran from April to August 2019. More than 120 teams participated and it attracted the interest of researchers from both communities --as evidenced from the top-5 ranked teams and the solutions they submitted. As such, we consider the secondary objective of the Challenge has been met successfully.

But what of the main goal of the Challenge, i.e. automating the extraction of useful parameters from transiting exoplanet light curves in the presence of stellar spots? A large number of solutions achieved the desired precision of 10\,ppm in photometric flux for correctly predicting the relative transit depth per each wavelength from the noisy light curves.

 The solutions of the top-5 ranking teams that participated in the Challenge are presented in detail in this paper. Most solutions amount to constructing highly non-linear (w.r.t. the raw data) models with minimal pre-processing using deep neural networks and/or ensemble learning methods\footnote{Ensemble methods are machine learning algorithms that construct powerful predictive models by combining multiple weaker predictors~\citep{polikar2006ensemble}.}. As we will see however, there exist comparably good --in terms of the precision of the obtained predictions-- approaches that involve obtaining meaningful (i.e. informed by physics) statistics from the light curves and then training models that are linear w.r.t. them. 

Just like the Challenge itself, this paper also intends to serve a dual purpose. Its primary goal is to describe the research problem of obtaining good predictions of the relative transit depth per each wavelength from simulated Ariel-like light curves distorted by photon noise and stellar spot noise, along with the solutions provided by the Challenge's winners and their implications. Its secondary aim is to promote interaction between exoplanetary scientists and machine learning researchers. As such it is written in a language accessible to both audiences and --we hope-- it contains useful information for exoplanetary scientists wishing to organise their own machine learning challenge.

\section{Exoplanet Background}

Due to the interdisciplinary nature of this article, we here provide a very brief high-level introduction to transmission spectroscopy of exoplanets. Readers familiar with the field can safely skip this section, for a more in-depth review to exoplanetary spectroscopy we refer the reader to the relevant literature \citep[e.g.][]{madhu19,tinetti12,sharp07}.

\begin{figure}
    \centering
    \includegraphics[width=1.0\textwidth]{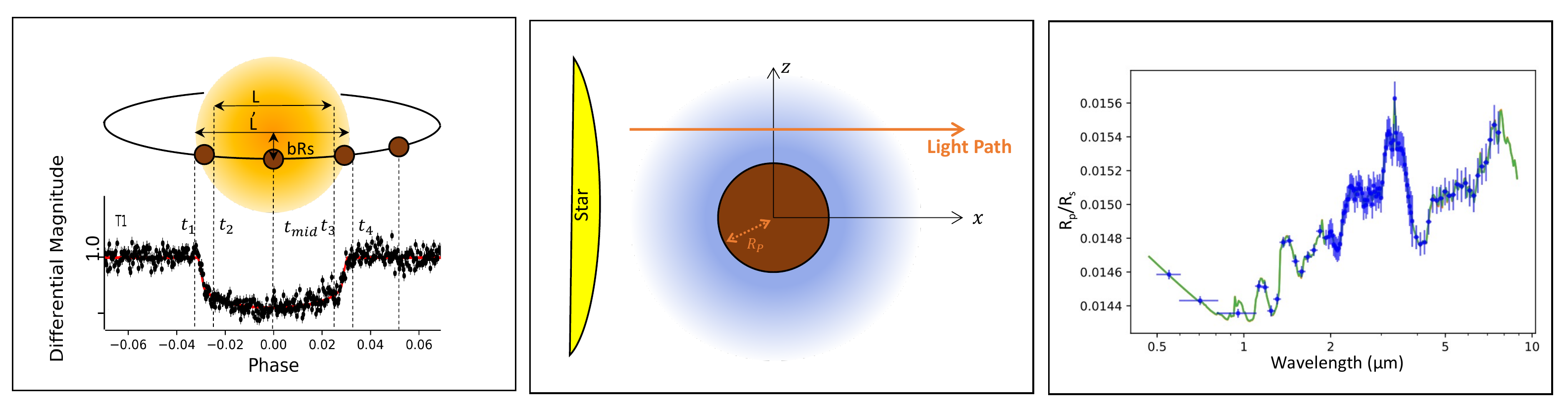}
    \caption{Left: Schematic of an exoplanet transit. The planet passes in-front of the star, obscuring some of the star's light. This leads to a characteristic dip in stellar flux observed as a function of orbital phase. Middle: Schematic view of transmission spectroscopy whereby some of the stellar light `shines through' the gaseous envelope of a planet. Right: A simulated transmission spectrum of the Ariel mission. Blue are the observed data points and green is a theoretical atmospheric model. Figures courtesy of C. Changeat and adapted from \citet{changeat20a}.  }
    \label{fig:transit_example}
\end{figure}

When a planet orbits its host-star in our line of sight, we will observe a regular dimming of the stellar flux when the planet passes between us and the host-star. This is referred to as a transit event. Similarly, when the planet is eclipsed by the host star, we will observe a small dip due to the loss of the planet's thermal or reflected light. In Figure\,\ref{fig:transit_example}a, we show a schematic view of a transit and the resulting dip in the stellar flux time-series, also known as a `lightcurve'. The depth of this lightcurve, $D$, is typically of the order of 1$\%$ for a Jupiter sized planet and a Sun like star. To first order, this dip can be described by the ratio of the planet to stellar radius, $D = R_p/R_\ast$ (also referred to as `relative radius'). For an in depth explanation of the transit geometry, see \citet[e.g.][]{seager03}. 

When a planet harbours an atmosphere, some of the stellar light will `shine through' the planet's gaseous envelope (Figure\,\ref{fig:transit_example}b). Depending on the atmospheric composition some light will be absorbed and/or scattered at specific wavelengths of light by the atmospheric gases, clouds and aerosols. This leads to a wavelength dependent `loss' of stellar flux observed, which is equivalent to a  perceived increase in planetary radius from an observational viewpoint. An accuracy of ca. 1 in 10$^4$ in flux measurements is typically required for a Jupiter size planet to observe this effect. Figure\,\ref{fig:transit_example}c is a simulation of the resulting transmission spectrum of a hot-Jupiter planet as observed by the Ariel Space Mission. The transmission spectrum includes absorption signatures of H$_2$O, CH$_4$, CO as well as Rayleigh scattering and collision induced absorption by hydrogen and helium \citep[][fig. 1]{changeat20a}.

\section{The Challenge}

\subsection{Data Generation}\label{sec:problem_statement}

For the purposes of the Challenge, we used the Ariel target-list produced by \cite{target_list} to generate simulated light curves for all the $2097$ planets in the list. For every planet we produced 55 light curves, one for each wavelength channel corresponding to Ariel Tier 2 resolution (between 0.5 and 8.0 $\mu$m). In addition, all the light curves covered observations of 5 hours, centred around the transit, with a time step of one minute. The uniformity in wavelength and time resolution is not realistic, but only used to make the dataset accessible to all the participants without the need of renormalisation (which would require knowledge on the transit modelling).

The simulated light curves were computed as follows:
\begin{enumerate}

\item As a first step we calculated the limb-darkening coefficients (using the quadratic law) for every host star in the target list and for every wavelength channel. We used the EXOTETHYS package \citep{exotethys} and the stellar parameters for temperature and gravity provided in the target list, assuming zero metallicity for all the stars (the effect of metallicity is not strong). Also, we did not use the Ariel throughput as in this study we were only interested in the narrow wavelength channels, and any intra-channel variations due to the Ariel throughput are minimal.

\item We then calculated the planet-to-star radius ratio, $R_\mathrm{p}/R_*$, for every planet in the target list and for every wavelength channel. This calculation was made using the TauRex atmospheric retrieval framework for exoplanets \cite{taurex} and the planet parameters for temperature, mass and radius (all provided in the target list), assuming the presence of water vapour and methane in the atmosphere with abundances that varied uniformly at random from planet to planet between 0.001\% and 0.1\%. The values for the abundances were an arbitrary choice, as the scope of using a spectrum was only to include some variability, of any kind, in the $R_\mathrm{p}/R_*$ parameter from one wavelength channel to another.

\item{The next step was to define the spot model parameters for every host star in the target list. These parameters were: 
\begin{itemize}
\item Spot coverage: This parameter corresponds to the percentage of the stellar surface that is covered by spots. We set this parameter to 10\% for every host star in the target list. In reality this parameter decreases with stellar temperature and initially we incorporated this in the model. However, it became clear that in the more realistic case, the number of spots that influence the light curves is very small, leading to almost noise-free data. For this reason, we chose to use the fixed value of 10\% in order to have a stronger spot effect on our data. This choice resulted in a simulated dataset with many more spot-crossing events than in a real dataset, suitable for the purposes of the challenge. 
\item Spot temperature: This parameter corresponds to the effective temperature of the spots, which is naturally lower that the effective temperature of the star. We calculated this parameter for every host star in the target list as a function of its temperature ($T_*$, provided in the target list), as described in \cite{subi_thesis}, adjusted from \cite{spot_temp}:
\begin{equation}
T_\mathrm{spot} = T_* - (0.0001343 \times T_*^2 - 0.6849 \times T_* + 1180.0)
\end{equation}
\item Spot contrast: This corresponds to the contrast between the brightness of the stellar surface and the brightness of the spots. We calculated this parameter for every host star in the target list and for every wavelength channel by integrating the respective PHOENIX stellar models \citep{phoenix} within each wavelength channel and dividing them.
\end{itemize}}

\item {Following the definition of the spot model parameters we created a set of spots for every host star in the target list. The spots were generated one by one, until the 10\% surface coverage was reached, and it was given three parameters:
\begin{itemize}
\item Latitude - uniformly at random generated number between -85 and 85 degrees
\item Longitude - uniformly at random generated number between 0 and 360 degrees
\item Angular diameter - randomly generated using a log-normal distribution, as described in \cite{subi_thesis}, based on  \cite{spot_size}:
\begin{equation}
\frac{dN}{dA} = M_A \exp{\left[-\frac{(\ln A - \ln \langle A \rangle)^2}{2 \ln \sigma_A}\right]}
\end{equation}

where $N$ is the number of spots, $A$ is the area of the spots, $M_A$ is the maximum of the distribution (adjusted to result in 10\% of total coverage), $\langle A \rangle = 0.62 \times 10^{-6} A_{1/2\odot}$ is the mean of the distribution, and $\sigma_A = 3.8 \times 10^{-6} A_{1/2\odot}$ is the standard deviation of the distribution.
\end{itemize}}

\item With the set of spots generated for each star in the target list, we used the KSint package \citep{ksint} to generate the spot-distorted light curves for every planet in the target list and for every wavelength channel. The input parameters for each light curve were: the set of spots, (number, position and dimensions of all the spots), the spot contrast parameter, the limb-darkening coefficients, the planet-to-star radius ratio, the stellar density (calculated from the stellar mass and radius provided in the target list), and the planet orbital parameters (period and inclination, provided in the target list) and a viewing angle to make sure that the transit happens at the middle of the observation. 

\item The final step was to add Gaussian noise to the light curve. No additional instrument systematics were assumed, as we aimed for the challenge to focus on correcting for the noise resulted from the stellar spots. The standard deviation of the Gaussian noise added was calculated from the overall noise on the transit depth estimation provided in the target list. This noise value depends on the stellar magnitude, the stellar temperature, the wavelength channel and the characteristics of the Ariel instrument. It is beyond the scope of this work to describe exactly how this level of noise is estimated. We refer the interested reader to \cite{target_list} for a detailed description.

\end{enumerate}

This process resulted in generating data for 2097 simulated observations, consisting of 55 light curves each (one per wavelength). We repeated the process 10 times with different instances of the spot set (step 4). This resulted in 20970 simulated observations consisting of 55 light curves each, distorted by stellar spots. Finally, for each instance of the spot set, 10 different instances of additive Gaussian photon noise were introduced (step 5). This resulted in 209700 simulated observations consisting of 55 light curves each, distorted by both stellar spot and photon noise. These 209700 simulated observations formed the final dataset of the Challenge. The different instances of the spot set were included to mimic multi-epoch observations, were the spot pattern is expected to change, while the different instances of additive Gaussian photon noise were included to mimic continuous observation were the spot pattern is not expected to change. Note that the two sources of noise (spots and Gaussian) are treated as independent. Most of the generated light curves only contained a single transit event, however a small number of them included planets with small enough orbital periods to allow for multiple transits\footnote{In case the light curve contained multiple transits, one of them was centered.}.

\begin{figure}
\centering
\subfloat[$0.7 \mu$m]{
  \includegraphics[width=0.35\textwidth]{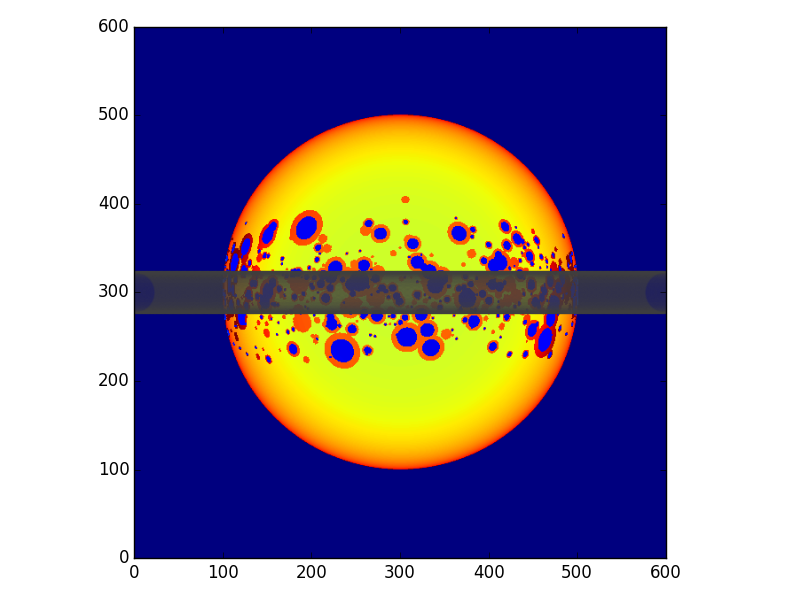}
}
\hspace{-10mm}
\subfloat[$5.6 \mu$m]{
  \includegraphics[width=0.35\textwidth]{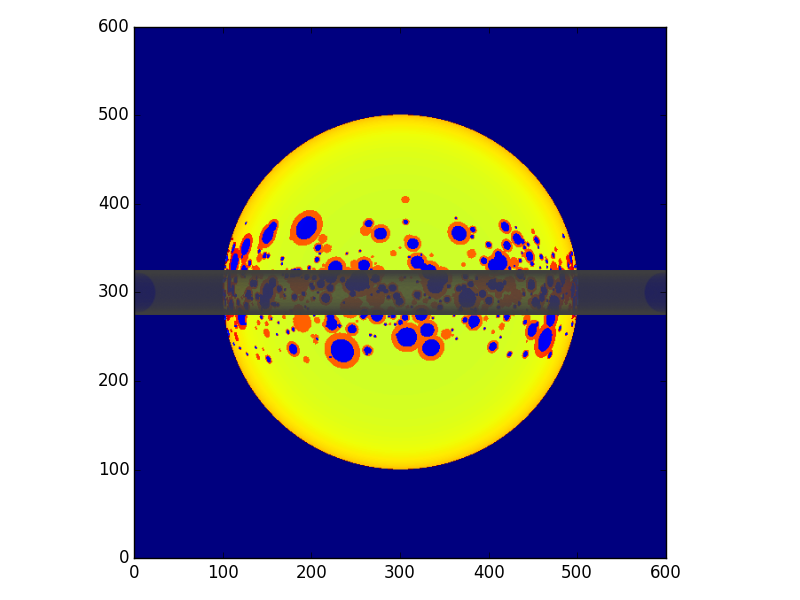}
}
\hspace{0mm}
\subfloat[$0.7 \mu$m]{
  \includegraphics[width=0.3\textwidth]{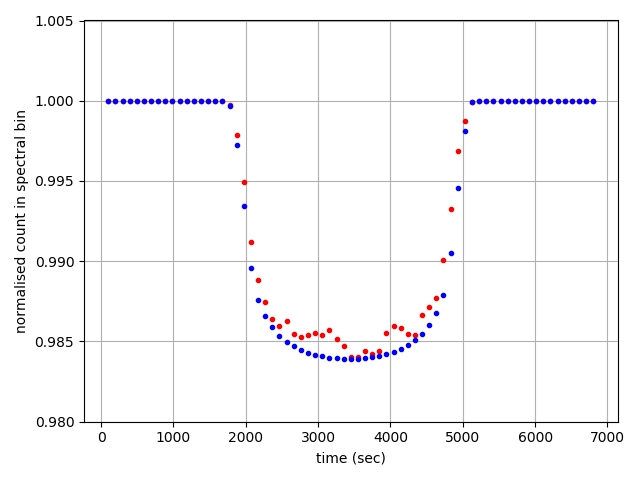}
}
\subfloat[$5.6 \mu$m]{
  \includegraphics[width=0.3\textwidth]{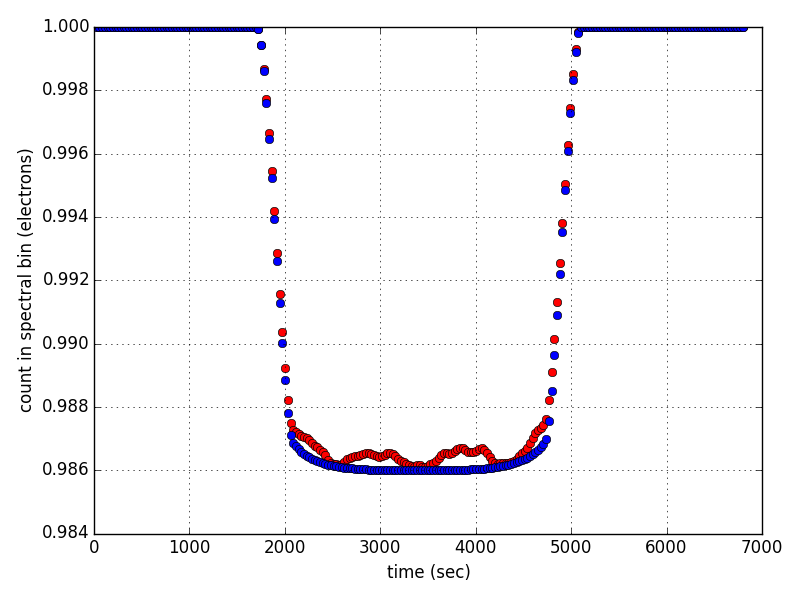}
}
\hspace{0mm}
\subfloat[$0.7 \mu$m]{
\includegraphics[width=0.3\textwidth]{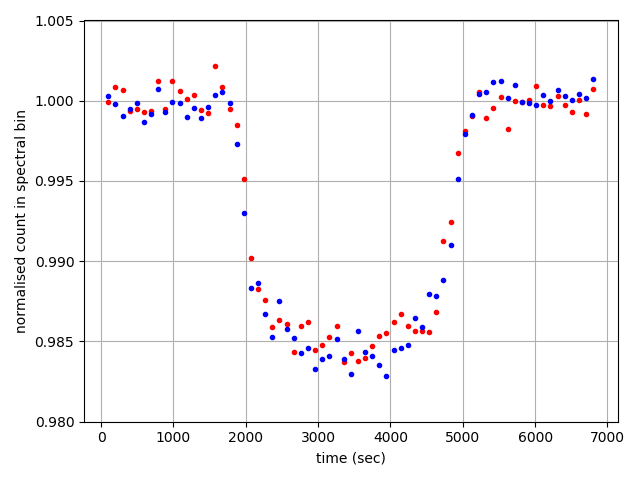}
}
\subfloat[$5.6 \mu$m]{
  \includegraphics[width=0.3\textwidth]{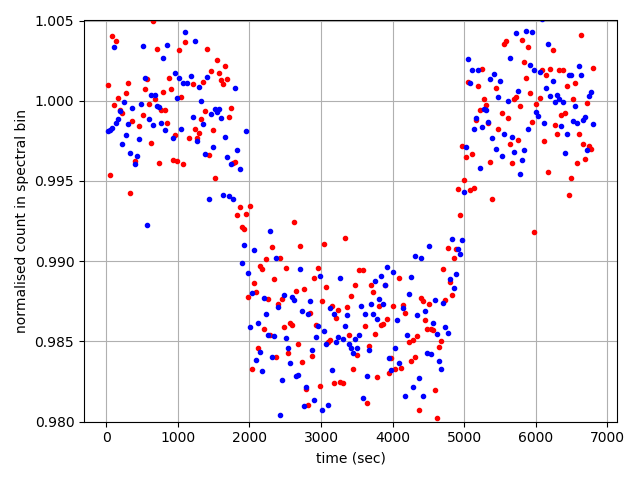}
}
\caption{Examples of simulations for two of the 55 wavelength channels, $0.7\,\mu$m and $5.6\,\mu$m. (a) \& (b), stellar surface simulations of a spotty star. Grey line shows the planet transit trajectory. The stellar surface limb brightness varies with wavelength. (c) \& (d) Normalised observed flux as the planet transits across the star without stellar photon noise. Blue shows the perfect transit across a spotless star; red shows the transit across a spotty star. (e) \& (f) same as (c) \& (d) but with stellar photon noise added.  \label{fig:sim}}
\end{figure}

Naturally, these details were unknown to the participants and neither were they used in the baseline solution. The aim of the Challenge was to infer the relative radii, either by explicitly modelling and subtracting, or by learning to ignore the photon and/or the stellar noise (or both). 
\subsection{Dataset Description \& Problem Statement}
Each \emph{datapoint} (a.k.a. an \emph{observation} or an \emph{example} in machine learning terminology) consists of a set of 55 noisy light curves (one per wavelength, corresponding to Ariel Tier 3 target resolution). Each light curve is a time series of 300 timesteps corresponding to 5 hours of observation by Ariel. We shall denote with ${x_{ij}}^{(t)}$ the relative flux at timestep $t\in[1,2,\dots,300]$ of the light curve at wavelength $j\in[1,2,\dots,55]$ of the $i$-th example. By ${\mathbf{x}_{ij}} = [{x_{ij}}^{(1)}, {x_{ij}}^{(2)}, \dots, {x_{ij}}^{(300)}]^\top$ we denote the entire light curve at wavelength $j$ of the $i$-th example. Finally, with ${\mathbf{X}_{i}} = [{\mathbf{x}_{i1}}, {\mathbf{x}_{i2}}, \dots, {\mathbf{x}_{i55}}]$ we denote all 55 light curves of the $i$-th example.

Along with the light curves, 
6 additional stellar and planetary parameters (all knowable in advance) were provided: the orbital period, stellar temperature, stellar surface gravity, stellar radius, stellar mass \& stellar $K$ magnitude. We shall denote these as ${z_{i1}}, {z_{i2}}, \dots, {z_{i6}}$, respectively, for the $i$-th example. Finally, with ${\mathbf{z}_{i}} = [{z_{i1}}, {z_{i2}}, \dots, {z_{i6}}]$ we shall refer to all the additional parameters of the $i$-th example collectively.

The noisy light curves and the 6 additional stellar and planetary parameters all constitute the quantities known in advance that we can use to alleviate the problem of stellar spots. In machine learning terminology they are the \emph{features (independent variables)} in our prediction task.

The goal is to construct a model that uses these to predict a set of 55 real values, the relative radii $[R_{p}/R_{*}]_{ij}$ (one per wavelength $j$, for any given datapoint $i$). In machine learning terminology this is a \emph{multi-target regression task}. The relative radii to be predicted are the \emph{targets (dependent variables)} of the multi-target regression problem. For convenience, we shall henceforth denote the relative radius at wavelength $j$ of the $i$-th example, $[R_{p}/R_{*}]_{ij}$, with ${y_{ij}}$. Finally, with ${\mathbf{y}_{i}} = [{y_{i1}}, {y_{i2}}, \dots, {y_{i55}}]$ we shall refer to all the relative radii of the $i$-th example collectively. Note the planet to host star relative radius $R_{p}/R_{*}$ is directly connected to the transit depth of the light curve, as the latter is equal to $\big(\frac{R_{p}}{R_{*}}\big)^2$. 




The value of the 55 targets is known only for the \emph{training examples} (the \emph{statistical sample}). The goal of the learning task is --ideally-- to construct a model $f({\mathbf{X}}, {\mathbf{z}}) = \mathbf{\hat{y}}$ such that $\mathbb{E}[L({\mathbf{y}} , \mathbf{\hat{y}})]$ is minimized, where $L({\mathbf{y}}, \mathbf{\hat{y}})$ denotes some measure of difference between the predictions $\mathbf{\hat{y}}$ and their corresponding true values ${\mathbf{y}}$ and $\mathbb{E}$ denotes expectation over the joint distribution of $\mathbf{X}, \mathbf{z}, \mathbf{y}$, i.e. --in statistical terminology-- the underlying \emph{population} from which the sample is drawn.

Once models are trained, they are evaluated on a separate \emph{test set}. The predictive performance of a model on a previously unseen test set (drawn from the same distribution as the training set), serves as a proxy for its performance in the population, the latter being intractable. The features of the test set examples
$\lbrace ({\mathbf{X}_i}, {\mathbf{z}_i}) | i\in{Test} \rbrace$
were provided to the participants and they had to upload their model’s predictions $\lbrace\mathbf{\hat{y}_i} | i\in{Test}\rbrace$ on them. The ground truth $\lbrace\mathbf{y}_i | i\in{Test}\rbrace$ for the test set examples was unknown to the participants in the duration of the Challenge. It was only used to produce a ranking score for their submitted solution, which we describe in the the next subsection.

\subsection{Evaluation}
\label{ssec:evaluation}

All datapoints generated for a uniformly random set of 1677 out of the 2097 of the total planets (i.e. about $80\%$ of the generated datapoints) were used as training data. All datapoints generated for the remaining 420 planets were used to form the test set (i.e. were only used for evaluation). That is, the training and test sets not only contained no datapoint in common, but they also contained no datapoint from the same planet in common.

After producing a model (i.e. a solution to the problem), the participants could upload the predictions of the model on the Challenge's website. Subsequently, this would assign a score on the model based on the quality of the predictions. The participants were ranked on a leaderboard on the basis of their best solution and the progress of each participant's solutions in time was tracked to inform them of the impact of each change they made on the resulting model's predictive performance. The leaderboard ranking determined the winners of the Challenge that would receive prizes (top 2 participants) and the top-5 participants whose solutions we will present in Section \ref{sec:solutions}.

The score assigned to each solution was a weighted average of the absolute error per target (i.e. on the relative radii) across all test set examples $i$ and all wavelengths $j$:
\begin{center}
\begin{equation}
\label{eq:score}
Score = 10^4 - \frac{\sum_{i \in Test} \sum_{j=1}^{55} w_{ij} 2 y_{ij} |\hat{y}_{ij} - y_{ij}|}{\sum_{i \in Test} \sum_{j=1}^{55} {w_{ij}}}10^6,
\end{equation}
\end{center}
where $y_{ij}$ is the true relative radius and 
$\hat{y}_{ij}$ the predicted relative radius of the $j$-th wavelength of the $i$-th test set example and the corresponding weight $w_{ij}$ is given by:
\begin{center}
\begin{equation}
w_{ij} = \frac{1}{ {{\sigma_{ij}}^2} {{\delta_{F_{ij}}}^2} },
\end{equation}
\end{center}
with ${\sigma_{ij}}^2$ being the variance of the relative stellar flux caused by the observing instrument at the $j$-th wavelength of the 
$i$-th example and ${\delta_{F_{ij}}}^2$ the variation of the relative stellar flux caused by stellar spots in the $j$-th wavelength of the 
$i$-th example. The value of $\sigma_{ij}$ is an estimation based on an Ariel-like instrument, given its current design, while $\delta_{F_{ij}}$ is calculated based on stellar flux $F_{ij}^{star}$ and the spot flux $F_{ij}^{spot}$ in the $j$-th wavelength of the $i$-th example:
\begin{center}
\begin{equation}
\delta_{F_{ij}} = 0.1\Bigg(1-\frac{F_{ij}^{spot}}{F_{ij}^{star}}\Bigg).
\end{equation}
\end{center}
As we see, both sources of noise (photon \& stellar spot) are wavelength-dependent and target-dependent (they depend on the star, therefore are different for each datapoint).

The higher the score, the better the solution's ranking. The maximum achievable score is 10000 (if $\hat{y}_{ij} = y_{ij}, \forall i,j$). The score is not lower-bounded (i.e. can be negative), but even naive `reasonable' models (e.g. predicting the average target value for all test datapoints) would not produce scores below 4000. 

The weights $w_{ij}$ of each target were unknown to the participants\footnote{For transparency of the evaluation process, the $w_{ij}$ coefficients of the test set examples, along with the ground truth (target values $y_{ij}$) became available after the end of the Challenge.}. A sensible strategy would thus be to try to predict all of them reasonably well. In other words, to train a model to minimize an unweighted loss $L({\mathbf{y}}_i , \mathbf{\hat{y}}_i)$ like the Mean Squared Error (MSE), $L({\mathbf{y}}_i, \mathbf{\hat{y}}_i) = ({\mathbf{\hat{y}}_i -\mathbf{y}}_i)^2$, the Mean Absolute Error (MAE),
$L({\mathbf{y}}_i, \mathbf{\hat{y}}_i) = |\mathbf{\hat{y}}_i -{\mathbf{y}}_i|$,
or their relative error counterparts: $L({\mathbf{y}}_i, \mathbf{\hat{y}}_i) = \Big(\frac{{\mathbf{\hat{y}}_i -\mathbf{y}}_i}{\mathbf{\hat{y}}_i}\Big)^2$ or  $L({\mathbf{y}}_i, \mathbf{\hat{y}}_i) = \frac{|{\mathbf{\hat{y}}_i -\mathbf{y}}_i|}{\mathbf{\hat{y}}_i}$, respectively. Indeed, this is the approach taken by the top-5 participants and in training the baseline model.

\subsection{Rules, Logistics \& Organization}

To allow for the broadest possible participation, the set of rules of the Challenge was the minimal possible. There was no restriction on the models, algorithms or data preprocessing techniques, neither on the programming languages, environments or tools used for their implementation. The participants were also free to use data augmentation techniques, pretrained models or any prior domain knowledge not included in the provided dataset. Finally, they were free to choose their own way of splitting the training data between training and validation sets. 

The participants were limited to 1 submission every 24 hours. This was a measure taken to limit traffic on our website and --most crucially-- to prevent the extend to which the solutions would be overfitting to the test set. Indeed, although the test set contains previously unseen examples by the model and the participants could not have access to the ground truth itself, the presence of a leaderboard is effectively causing some \emph{information leakage from the test set}. Simply put, just adapting the strategies to the ranking score signal, participants could increase their scores by effectively overfitting on the particular test set. Limiting the number of daily submissions alleviated this effect. In retrospect, an even stronger strategy to prevent this would have been to only use part of the test set to produce the leaderboard ranking score during the Challenge and only use the full test set to produce the final ranking after the Challenge closes. In future machine learning challenges we will adopt this evaluation scheme. For now we should keep in mind that small differences in the ranking scores of solutions presented in Section \ref{sec:solutions} are not necessarily indicative of true generalization (i.e. ability to predict well on new examples). 

The participants were allowed to form teams, provided they participated in only one entry. The remaining rules handled how prizes would be split among teams, how ties would be handled and ensuring that any winning entry would have to beat the baseline model.

\subsection{Description of  Solutions}

To facilitate comparisons among the solutions discussed in the paper and to demonstrate the typical steps of training and evaluating models using machine learning methodology, we split the description of the solutions into 3 parts:  (i) preprocessing, (ii) model / architecture, (iii) training / optimization.

The `preprocessing' part will describe any transformation of the raw data (either in terms of features or of observations) before giving it as input to a learning algorithm. The `model' part is concerned with the general \emph{class of models} (i.e. their \emph{parametric form}) which the learning algorithm is exploring (e.g. deep neural networks of a given \emph{architecture}\footnote{By the term ``architecture" we collectively refer to the number, type and connectivity of the neurons comprising a neural network.}, random forests of 10 trees of maximal depth 5, linear models of the form $y = ax_1 + bx_2+c$). Finally, the `training' part is concerned with the specifics of the optimization of the \emph{parameters} of the model (i.e. the weights of the neural network, the derivation of the decision trees or the inference of the linear coefficients $a,b,c$ in the examples below). It covers the \emph{hyperparameters} used in the learning/optimization algorithm, along with the loss function it minimizes and the final evaluation method.

Wherever necessary, we will clarify the purposes behind modelling choices or training methodologies in all solutions described. However, a detailed treatment of models 
like deep neural networks (DNNs) is beyond the scope of this paper. We direct the interested reader to \cite{DLGoodfellow} and \cite{DLKeras}.

\subsection{Baseline Solution}
As a baseline solution, we trained a fully connected DNN\footnote{Fully connected DNNs are the earliest and most popular type of DNN architecture. They are also known as multi-layer perceptrons (MLPs) or `dense' neural networks.} on a sample of 5000 training examples selected uniformly at random. The neural network uses all 55 noisy light curves, $\mathbf{X}_i$ to predict the 55 relative radii directly. It does not make use of any of the additional stellar \& planetary parameters $\mathbf{z}_i$.

\subsubsection{Preprocessing}
The noisy light curves have undergone the following preprocessing steps:

i) Each light curve was smoothed using a moving median of window 3 (i.e. each value replaced by the median of itself and its two adjacent values). This was done to remove flux values that are obvious outliers.

ii) In any light curve, any value (relative flux) that was above 1 was clipped to 1. This was done because the maximal relative flux during transit is 1.

iii) All values were normalized for the transit depths to lie roughly within the range $[0,1]$. Doing so allows for faster and more stable training of models like DNNs. The normalization was carried out per wavelength and was performed as follows:

First, we computed the average transit depths per wavelength from the target values $\bar{y}_{j}$ on a sample of 10000 random training examples. For every wavelength $j$, we then applied the transformation:
 \begin{equation*}
 x_{ij}^{(t)} \leftarrow (x_{ij}^{(t)} - (1-2\bar{y}_{j}^2)) / 2\bar{y}_{j}^2.
 \end{equation*}
This was done to have the maximal relative flux values at exactly 1 and the transit depths around 0, leveraging the fact that the transit depths of the light curves are the squares of the relative radii (targets).

\subsubsection{Model/Architecture}
We used a fully connected DNN with 5 2D-hidden layers, all of which consisted of 1024 units $\times$ 55 channels, the j-th channel receiving as input the light curve $\mathbf{x}_{ij}$ for each example. After these, we added a flattening layer followed by a linear layer of 55 outputs, the j-th output corresponding to the predicted relative radius $\hat{y}_{ij}$ of each example. All other activation functions were rectified linear units (ReLUs). 

\subsubsection{Training/Optimization}
No batch normalization, regularization or dropout was applied in the training of the baseline model. The 5000 observations used were split into 4020 training and 980 validation examples (i.e. approximatelly 80\% training \& 20\% validation split) in such a way that the two sets contained no planets in common. The model was trained by minimizing the average MSE across all wavelengths using the Adam optimizer~\citep{kingma_adam:_2014} with a learning rate of $10^{-4}$ decaying with a rate of 0.01 and a batch size of 128. All remaining hyperparameters were set to default \texttt{Keras}\footnote{https://keras.io} values. The model was trained for a maximum number of 5 epochs without early stopping.

\section{Top-5 Solutions}
\label{sec:solutions}
By the end of the Challenge, 13 teams had beaten the score attained by the baseline solution we just presented. In this section, we will present the top-5 ranked solutions. Their relative ranking in the final leaderboard and scores they achieved under Eq.(\ref{eq:score}) are shown in Table \ref{table:leaderboard}.

\begin{table}[h]
\centering
\begin{tabular}{c|c|c}
Rank & Team         & Score \\ \hline
1    & SpaceMeerkat & 9813  \\
2    & Major Tom    & 9812  \\
3    & BV Labs      & 9808  \\
4    & IWF-KNOW     & 9805  \\
5    & TU Dortmund University    & 9795  \\
14   & Baseline     & 8726 
\end{tabular}
\caption{Final leaderboard showing rank \& score under Eq.(\ref{eq:score}) achieved by each of the top-5 entries and the baseline. } \label{table:leaderboard}
\end{table}

\subsection{SpaceMeerkat's solution}

\textit{SpaceMeerkat} is comprised of James M. Dawson, an Astrophysics PhD student at Cardiff University. \textit{SpaceMeerkat}'s solution is a 1D-CNN, designed to retain architectural simplicity, while exploiting the power of GPU accelerated machine learning. The largest gain in the model's predictive power came from the extensive testing of different prepossessing operations.

\subsubsection{Preprocessing}
The data was split into 80\% training and 20\% test sets. In order to remove outlier flux values in the raw light curves, an initial smoothing was conducted on each time series $\mathbf{x}_{ij}$. The mean flux value in each non-overlapping bin of width 5 was calculated in-place along each time series leaving each observation $\mathbf{X_{i}}$ as a smoothed multi-channel array of dimensions $60 \times 55$. 
A normalisation operation was performed on the training set prior to its use for training machine learning models. For each of the 55 wavelengths, the medians across all datapoints of the lowest 1\% of flux values in each light curve for a given wavelength were calculated. These 55 percentile medians (henceforth `median offsets') are therefore equal to 
\begin{equation}
\mathbf{\kappa}_{j} = \text{med}\{\mathbf{P}_{1\%}(\mathbf{x}_{ij}^{(t')})\},
\end{equation}
where $\mathbf{P}_{1\%}(\mathbf{x}_{ij}^{(t')})$ denotes the 1st percentile of the set of all flux values $x_{ij}^{(t')}$, $t' \in \lbrace 1, 2, \dots, 60 \rbrace$ for a given datapoint $i$ and wavelength $j$, and $med\lbrace \cdot \rbrace$ denotes median across all datapoints $i$. The light curves were then divided by 1 minus the median offsets and the resulting flux values were thus
\begin{equation*}
x_{ij}^{(t')} \leftarrow x_{ij}^{(t')}/(1-\mathbf{\kappa}_j).
\end{equation*}
This normalisation allowed the data to lie roughly within the range $[0, 1]$ but with leniency for allowing the existence of extremely shallow or deep transits. Any remaining flux values above the normalisation range were clipped to 1. This was done to encourage the model to focus on the lower flux valued regions where most of the transit-depth information lies. The preprocessing of light curves makes use of \texttt{Astropy}\footnote{http://www.astropy.org}, a community-developed Python package for Astronomy \citep{astropy_collaboration_astropy:_2013,astropy_collaboration_astropy_2018}.

\subsubsection{Model/Architecture}

The model used in this solution is a convolutional neural network (CNN)~\citep{lecun1995convolutional}\footnote{CNNs are designed to excel in tasks in which translational invariance is important, i.e. we are looking for particular patterns anywhere in the input data. As such, they are especially popular in image-based tasks. However, they are very successful even outside this setting, as they effectively reduce the number of trainable parameters of a neural network (compared to a feedforward DNN of the same depth). This means they are more computationally efficient to train and more resistant to overfitting.}. The data is presented to the CNN as a 1D vector and 1D convolutions \& pooling operations are applied in order to maintain a principled simplicity to the final solution. The architecture of the CNN is shown in Table \ref{table:spacemeerkat_architecture}. The model was built using \texttt{PyTorch 0.4.1}\footnote{\url{http://pytorch.org/}}, an open source machine learning framework for Python users. 
The output of layer `Lc5' in Table \ref{table:spacemeerkat_architecture} is concatenated with the additional stellar \& planetary parameters: the orbital period, stellar surface gravity, stellar radius, stellar mass \& stellar $K$ magnitude, i.e. $[{z_{i1}}, {z_{i3}}, \dots, {z_{i6}}]$ for each example, to form the 1D linear input for layer `Lc6'. The additional parameters did not undergo any normalisation and were presented to the network in their raw form.

\begin{table}
\centering
\begin{tabular}{llll}
\hline
Name   & Layer/Operation & Dimensions      & Filter \\ \hline
Input  & None            & (256,1,1,3300)   & None   \\
Conv1  & 1D convolution  & (256,32,1,3300)  & (1,3)   \\
ReLU   & ReLU            & None            & None  \\
AP1    & 1D average pool & (256,32,1,1650)  & (1,2)   \\
Conv2  & 1D convolution  & (256,64,1,1650)  & (1,3)  \\
ReLU   & ReLU            & None            & None  \\
AP1    & 1D average pool & (256,64,1,825)   & (1,2)   \\
Conv3  & 1D convolution  & (256,128,1,825)  & (1,3)  \\
ReLU   & ReLU            & None            & None  \\
AP1    & 1D average pool & (256,128,1,275)  & (1,2)   \\
Lc1    & Linear          & (256,1,1, 35200) & None  \\
ReLU   & ReLU            & None            & None   \\
Lc2    & Linear          & (256,1,1, 2048)  & None   \\
ReLU   & ReLU            & None            & None   \\
Lc3    & Linear          & (256,1,1, 1024)  & None   \\
ReLU   & ReLU            & None            & None   \\
Lc4    & Linear          & (256,1,1, 512)   & None   \\
ReLU   & ReLU            & None            & None   \\
Lc5    & Linear          & (256,1,1, 256)   & None   \\
ReLU   & ReLU            & None            & None   \\
Lc6    & Linear          & (256,1,1, 60)    & None   \\
Output & None            & (256,1,1,55)     & None   \\ \hline
\end{tabular}
\caption{The CNN architecture used in the solution by the SpaceMeerkat team (Ranked 1st). The table follows the standard \texttt{PyTorch} format. The 1st column lists the name of each layer/operation, the 2nd column its type, the 3rd the dimensions of its output tensors (hence inputs to the next layer). These follow the convention (batch size, number of channels, height, width). The filter column shows the dimensions (height, width) of kernels used to perform the convolution and pooling operations. Layer `Lc6' is notable as this is where the additional planetary parameters $\mathbf{z}$ are introduced into the network.}
\label{table:spacemeerkat_architecture}
\end{table}

\subsubsection{Training/Optimization}

The CNN was trained for 75 epochs (i.e. was presented with the entire training set 75 times), on a single NVIDIA TITAN Xp GPU. The model was trained using batches of 256 examples. Rather than presenting the CNN with examples of dimensions $60 \times 55$ (as generated by the preprocessing step), each example was flattened into a single vector of length $3300$. Initial investigation showed that 1D convolutions over the flattened inputs produced significantly better results than 2D convolutions over the 2D preprocessed inputs. The model was trained by minimizing the MSE loss (see $\S$\ref{ssec:evaluation}) using the standard Adam optimiser and an initial learning rate of $1\times10^{-3}$ decaying by $10\%$ the existing rate, every epoch. No early stopping was used, as we observed no increase of the validation error during training to indicate the presence of overfitting. No additional form of regularization (e.g. batch normalisation, dropout, or explicit regularisation) was used in the training procedure. All remaining hyperparameters were set to default \texttt{PyTorch} values. The code for this solution is publicly available on GitHub\footnote{Solution by \emph{SpaceMeerkat} (Ranked 1st): \url{https://github.com/SpaceMeerkat/ARIEL-ML-Challenge}}.

\subsection{Major Tom's solution}
Major Tom took second place on the ARIEL ML challenge scoreboard. The team composed of machine learning researchers from the Data Science Research and Analytics (DSAR) group at the University of Tuebingen (Germany).  The goal of the team's solution is to provide an easy to use ML tool, with minimal data preprocessing effort and a fast inference step. The result is a fully-integrated deep learning solution whose final code is publicly available online\footnote{Solution by \emph{Major Tom} (Ranked 2nd): \url{https://github.com/unnir/Ariel-Space-Mission-Machine-Learning-Challenge}}.


\subsubsection{Preprocessing}
The main motivation behind this solution was to create a robust statistical model that can handle outliers and noisy data. Therefore, we deliberately do not apply any heavy preprocessing to the data beyond the rescaling of the features and the targets. Since all measurements in time series $\mathbf{x}_{ij}$ are mostly distributed around 1 (see, for example Figure~\ref{fig:bvl_denoising}), we used the following rescaling of the data, in order to emphasise the differences between measurements: 
\begin{equation*}
   x_{ij}^{(t)} \leftarrow (x_{ij}^{(t)} - 1) \times 1000.
   \label{eq:data_transform}
\end{equation*}
We apply a similar transformation technique to the target variable $y$: 
\begin{equation*}
    y_{j} \leftarrow  y_{j} \times 1000.
\label{eq:target_transform}
\end{equation*}

\subsubsection{Model/Architecture}

\begin{figure}
  \centering
  \includegraphics[scale=0.68]{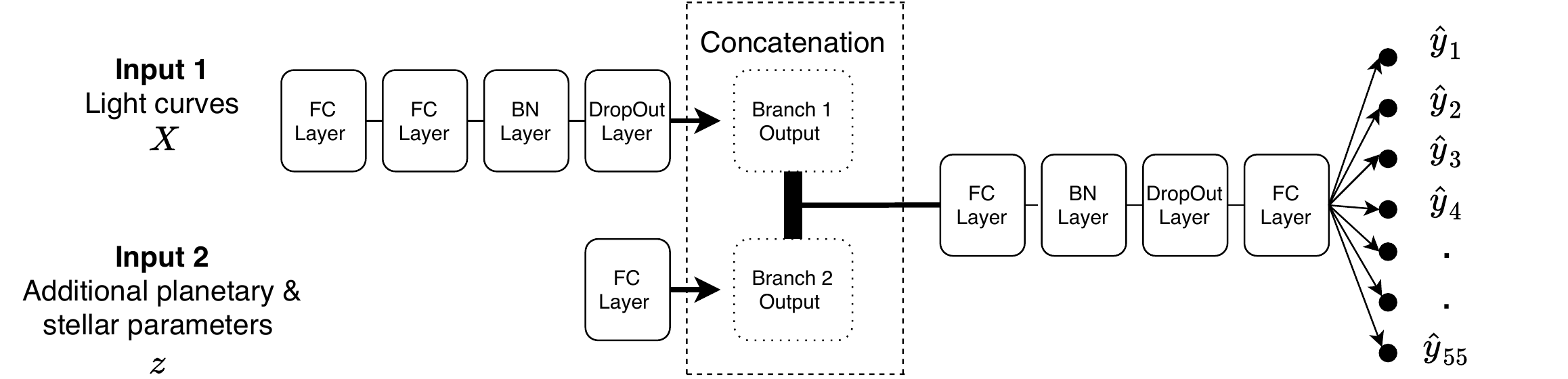}
  \caption{The deep learning model architecture proposed by the Major Tom team (Ranked 2nd). The model has two separate inputs: one for the measurements $\mathbf{X}_i$, the second for the additional stellar \& planetary parameters $\mathbf{z}_i$. The two branches are subsequently concatenated and higher level non-linear features combining information from both are extracted.}
  \label{fig:majortom_architecture}
\end{figure}

 We used a multiple-input and multiple-output DNN model with fully-connected (FC), Batch Normalization (BN) \citep{ioffe2015batch}, and Dropout \citep{srivastava2014dropout} layers\footnote{Both batch normalization and dropout are commonly used techniques to prevent overfitting in neural networks.}. The final architecture is presented in Figure \ref{fig:majortom_architecture}. It consists of two separate branches. The first branch uses as input the light curves $\mathbf{X}_i$, and the second, the additional stellar \& planetary parameters $\mathbf{z}_i$. After several non-linear transformations, the outputs of the two branches are concatenated into one and higher level non-linear features combining information from both are extracted. The output layer has 55 neurons, the $j$-th neuron mapping to the (rescaled) predicted relative radius $y_{ij}$ of a given example. We utilized exponential linear unit (ELU) activations in all but the last two layers, where ReLUs and linear activation functions are used, respectively.

\subsubsection{Training/Inference}
We train the DNN using the \textit{NAdam} optimization algorithm \citep{dozat2016incorporating}
and a \textit{cyclic learning rate} as described in \cite{smith2017cyclical}. The number of epochs was set to 1000, and the batch size to 3048. We selected the MSE as the loss function. We train the proposed model using 10-fold Cross-Validation with early stopping based on the validation loss with the patience equals to $20$.
The neural network was implemented using the Keras/Tensorflow deep learning framework (\cite{tensorflow2015-whitepaper}). The entire training step took $\approx$ 30 hours using a single NVIDIA P100 GPU.

For the inference step, we used an ensemble consisting of all 10 models produced in the cross-validation steps; the final prediction is the average of all estimates from the 10 models.  


\subsection{BVLabs' solution}
The team BVLabs took third place in the challenge. It is comprised of researchers and data scientists from the Jo\v{z}ef Stefan Institute and Bias Variance Labs. The team's solution relied on denoising the input data, the use of tree ensembles and fully-connected neural networks.

\subsubsection{Preprocessing}
\begin{figure}
    \centering
    \begin{tabular}{cc}
    \includegraphics[width=0.47\textwidth]{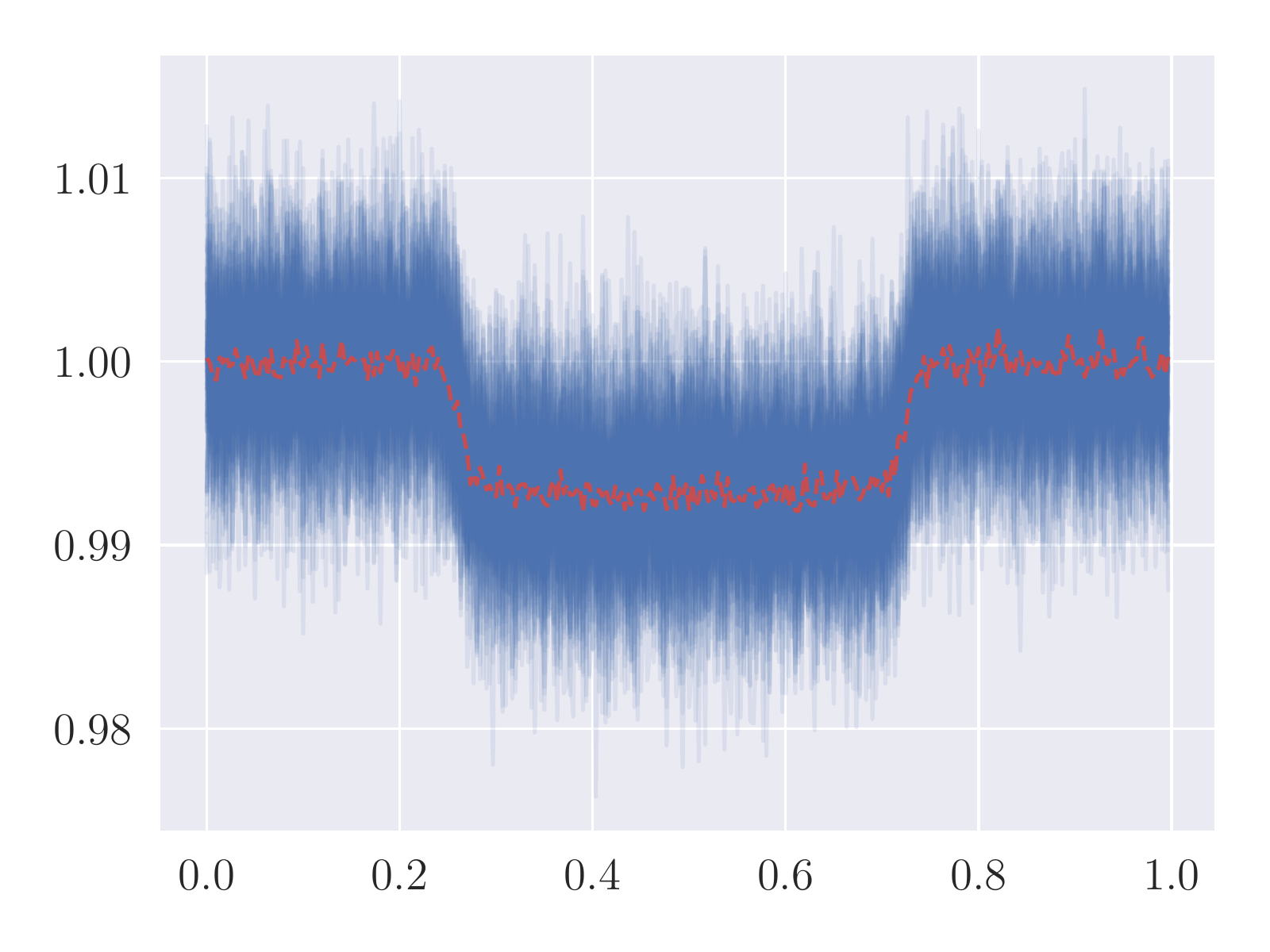} &
    \includegraphics[width=0.47\textwidth]{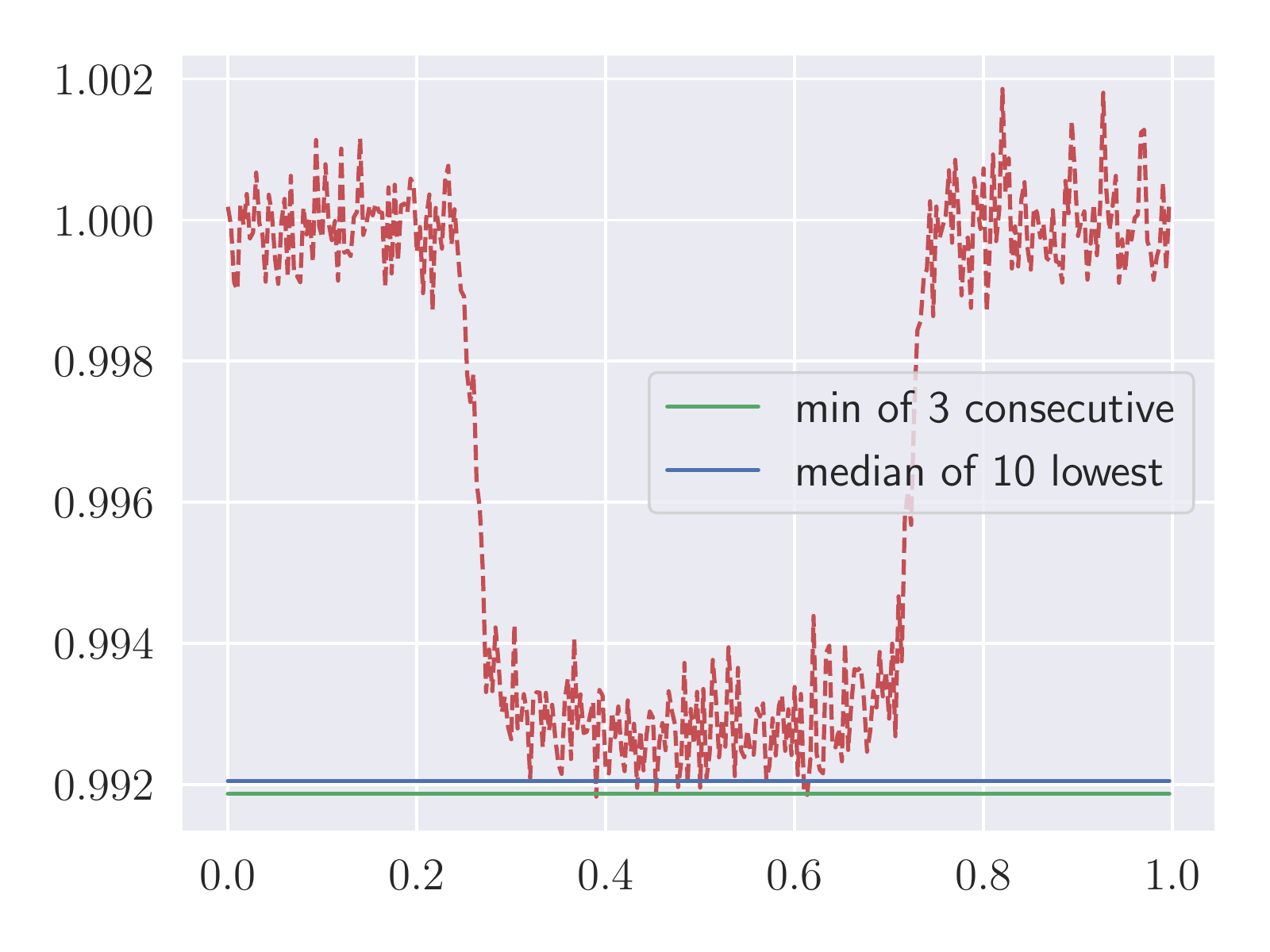} \\
    \hspace{10mm} (a) & \hspace{10mm}  (b)
    \end{tabular}
    \caption{Preprocessing of the light curves by the BV Labs Team (Ranked 3rd). Image (a) shows the light curve before (blue) and after (red) noise instance aggregation. Image (b) shows the features extracted from the denoised data. Both images show the data for star-planet pair 113, channel 25.}
    \label{fig:bvl_denoising}
\end{figure}

For each star-planet pair, we have 10 stellar spot noise instances and for each  stellar spot noise instance we have 10 Gaussian noise instances. The data for each star-planet pair can therefore be represented as a tensor with dimensions $(10, 10, 55, 300)$. For a fixed stellar spot noise instance, we computed the element-wise mean flux matrix over the 10 Gaussian noise instances which decreases the noise in the data. This can be seen as aggregating multiple measurements of the same target to decrease the variance of the observation. We are left with tensors with dimensions $(10, 55, 300)$. Next, we compute element-wise medians over the 10 stellar spot noise instances, leaving us with tensors with dimensions $(55, 300)$. An example of the result of this denoising process is presented in Figure~\ref{fig:bvl_denoising}a. 

The maximum flux (without noise) is always 1, whereas the minimal flux gives information about the planet radius. To further compensate for the noise, we do not use the minimal flux directly. Instead, we calculate two values: the minimum of the average of 3 consecutive flux values, and the median of the 10 lowest flux values. An example of the extracted values is shown in Figure~\ref{fig:bvl_denoising}b.

We also estimated the amount of energy that stars emit at operating wavelengths of the ARIEL spacecraft. \citet{tinetti} list the 5 operating ranges of ARIEL. We divided each range into 11 bins of equal length, to get the estimates of the 55 wavelengths. To calculate the energy at a given wavelength, we used Planck's law
$$	B(\lambda, T) \propto \frac{1}{\lambda^5} \frac{1}{\exp\left(\frac{h c}{\lambda k_B T}\right) - 1},$$
where $\lambda$ is the wave-length, $h$ is the Planck's constant, $k_B$ is the Bolzmann's constant and $c$ is the speed of light. The star temperature $T$ was one of the 6 stellar and planetary parameters (see Section~\ref{sec:problem_statement}).
In total we used 171 ($3\cdot55 + 6$) features: 3 features for each of the 55 channels (the 2 extracted from the flux values and the energy emitted) and the 6 stellar and planetary parameters.

\subsubsection{Model \& Training}
Our best performing model was a heterogeneous ensemble consisting of three models. The first model was a random forest of 500 trees \citep{breiman:rf}, as implemented in \texttt{scikit learn}\footnote{\url{https://scikit-learn.org/}}. The second model was an extreme gradient boosting \citep{friedman:boost} ensemble of 150 trees, as implemented in the \texttt{xgboost} library\footnote{\url{https://github.com/dmlc/xgboost}}. For both methods the parameters were optimized with cross-validation, and a separate model was learned for each channel. The third model was a multi-target (one model for all 55 channels) fully connected neural network with one hidden layer of 100 neurons. We used batch normalization, dropout (with a rate of 0.2) and ReLU activations. The network was optimized with the Adam optimizer for 1000 epochs, with a constant learning rate $10^{-3}$. As the loss function, average MSE across all targets was used. The network was implemented in \texttt{PyTorch}.

The weights of these 3 models in the final heterogeneous ensemble were optimized manually, with the best results obtained with a weight 0.15 assigned to the random forest, 0.25 to XGBoost and 0.6 to the neural network. The code is available online\footnote{Solution by \emph{BV Labs} (Ranked 3rd): \url{https://github.com/bvl-ariel/bvl-ariel}.}.

\subsection{IWF-KNOW's solution}

\begin{figure}
  \centering
  \includegraphics[width=0.8\textwidth]{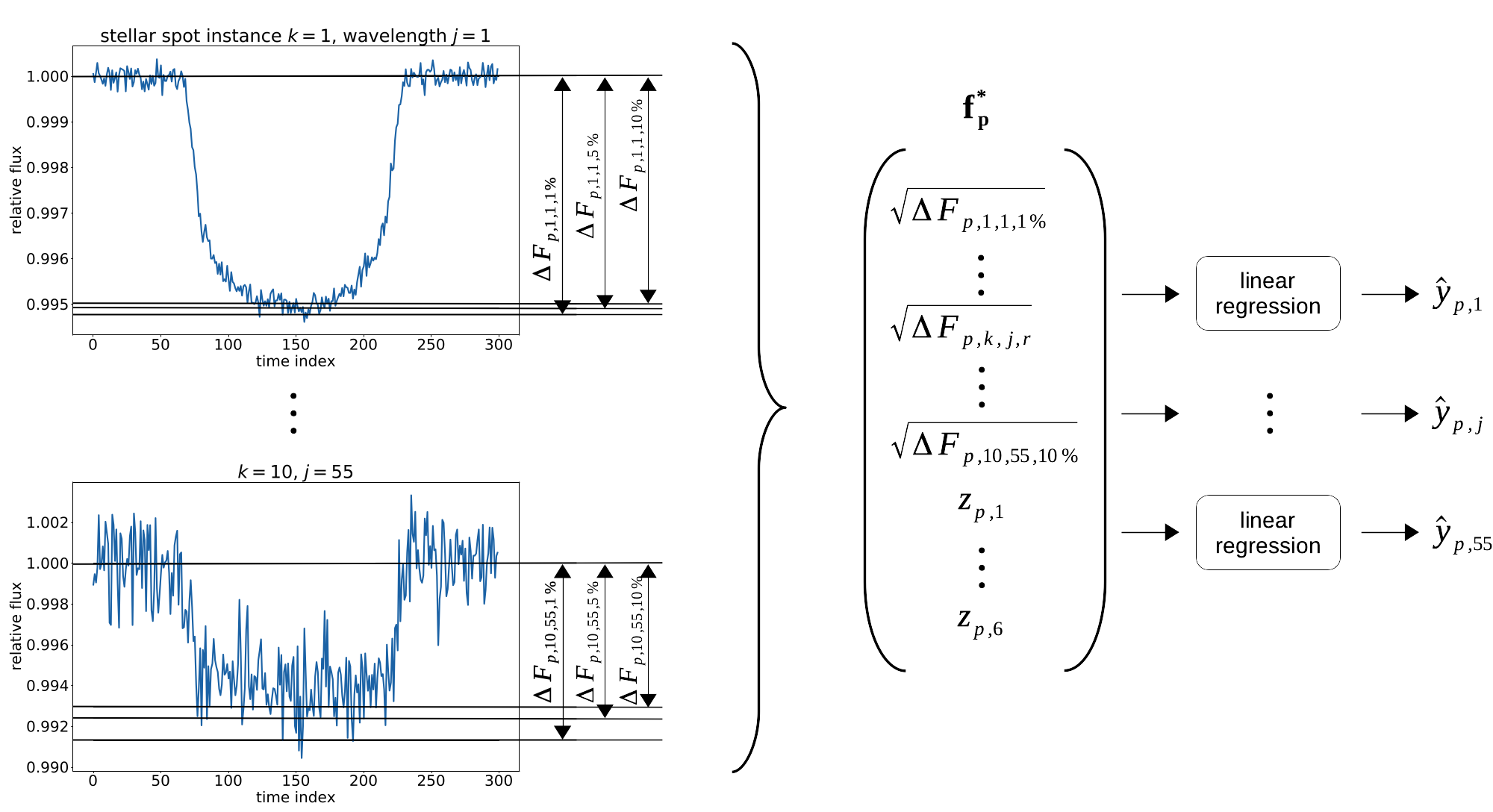}
  \caption{Regression pipeline of IWF-KNOW (ranked 4th). The lightcurves on the left are two examples in $\widetilde{\mathbf{X}}_{p,k}$. The minima of the light curves were estimated using the $1^{st}$, $5^{th}$, and $10^{th}$ percentiles. Subsequently, the minima were used to calculate the dips of the light curves $\Delta F_{p,k,j,r}$. The square root of all light curve dips $\Delta F_{p,k,j,r}$ belonging to the same planet $p$ (i.e. including all wavelengths $j$ and all stellar spot instances $k$), and additionally the stellar and planetary parameters $z_{p,1}, \dots, z_{p,6}$, were then gathered in the feature vector $\mathbf{f}_p^*$. The feature vector was z-score normalized (not shown in the graphic). Eventually, linear regressions were used to calculate the relative planet radius for each wavelength $j$.}
  \label{fig:iwfknow_pipeline}
\end{figure}

IWF-KNOW took the fourth place on the ARIEL ML challenge scoreboard, and comprised of researchers and data scientists from the \emph{Space Research Institute} (Austria), \emph{Know-Center} (Austria) and the \emph{University of Passau} (Germany). In contrast to the other top scorers who relied on deep learning approaches, their solution is based on a set of linear regressors, each of which is fast to train and easy to interpret (see Figure \ref{fig:iwfknow_pipeline}). The corresponding scripts can be found on Zenodo\footnote{Solution by \emph{KNOW-IWF} (Ranked 4rd) available under the DOI 10.5281/zenodo.3981141: \url{https://doi.org/10.5281/zenodo.3981141}.}.

\subsubsection{Preprocessing}

We re-indexed the examples $\mathbf{X}_i$, each of size $300 \times 55$, in a new matrix $\mathbf{X}_{p,k,l}$, where $p \in \lbrace 1,2,\dots, 2097 \rbrace$ indexes the planet, $k \in \lbrace 1,2,\dots, 10 \rbrace$ the stellar spot instance, and $l \in \lbrace 1,2,\dots, 10 \rbrace$ the photon noise instance. To reduce the photon noise, we averaged the examples $\mathbf{X}_{p,k,l}$ over the photon noise instances $l$ belonging to the same planet $p$ and stellar spot noise instance $k$, yielding the noise-reduced example matrix $\widetilde{\mathbf{X}}_{p,k} = \frac{1}{10} \sum_{l=1}^{10}{\mathbf{X}_{p,k,l}}$. $\widetilde{\mathbf{X}}_{p,k}$ was of size $300 \times 55$ and comprised of the light curves for each wavelength.
Subsequently, we calculated the differences between the maxima and minima of each light curve in $\widetilde{\mathbf{X}}_{p,k}$. The maxima were assumed to be $1$ as the light curves were already normalized, and the minima were estimated as the $1^{st}$, $5^{th}$, and $10^{th}$ percentiles. This yielded estimates $\Delta F_{p,k,j,r}$ of the dip of the relative light curve caused by a transit of planet $p$ for stellar spot noise instance $k$, wavelength $j$, and $r \in \lbrace 1\%, 5\%, 10\% \rbrace$ corresponding to the $1^{st}$, $5^{th}$, and $10^{th}$ percentiles. As the true dip $\Delta F_{p,j}$ of the relative light curve is approximately equal to the quadratic relative planet radius $\big(\frac{R_{p,j}}{R_{*,j}}\big)^2$, we took the square root of $\Delta F_{p,k,j,r}$ to obtain estimates of the relative planet radii:
$$\frac{R_{p,j}}{R_{*,j}} \approx \sqrt{\Delta F_{p,k,j,r}}$$

We then built a feature vector $\mathbf{f}_p$ comprised of the estimated relative planet radii belonging to planet $p$: $$\mathbf{f}_p = \left[\sqrt{\Delta F_{p,1,1,1\%}}, \dots, \sqrt{\Delta F_{p,k,j,r}}, \dots, \sqrt{\Delta F_{p,10,55,10\%}}\right]$$
The feature vectors $\mathbf{f}_p$ were augmented by the stellar and planetary parameters provided. For that, we averaged the 6 stellar and planetary parameters $z_{i1}, z_{i2}, \dots, z_{i6}$ over all photon noise and stellar spot noise instances belonging to the same planet yielding $\mathbf{z}_p = [z_{p,1}, z_{p,2}, \dots, z_{p,6}]$. The averaged stellar and planetary parameters $\mathbf{z}_p$ were then appended to the feature vectors $\mathbf{f}_p$ yielding the augmented feature vectors $\mathbf{f}_p^*$. The length of $\mathbf{f}_p^*$ was $1656$, which resulted from 55 wavelengths, 3 percentile-based dip estimations, 10 spot noise instances, and 6 stellar and planetary features ($55 \times 3 \times 10 + 6$). Strictly speaking, the averaging was not necessary as the stellar and planetary parameters were the same for all instances of a planet (i.e.\ no noise was added to the stellar and planetary parameters).
Finally, the extended feature vectors $\mathbf{f}_p^*$ were z-score normalized\footnote{This type of normalization, also known as `standardization' is performed by subtracting for each feature of a given example the mean value of that feature across all examples and dividing by its standard deviation.}, separately for the training and test set, thus avoiding information leakage from the test set into the training set.

We also re-indexed the scalar targets $y_{i,j}$ in the training set as $y_{p,k,l,j}$. Subsequently, we aggregated targets by averaging over all stellar spot noise instances $k$ and photon noise instances $l$ belonging to the same planet $p$, yielding the targets $y_{p,j}$. However, averaging was again not strictly necessary as all photon noise and stellar spot noise instances of a planet had the same relative radius in the provided dataset.

\subsubsection{Model \& Training}
We set up a multiple linear regression model per wavelength $j$, resulting in 55 regression models:
$$ y_{p,j} = \beta_{0,j} + \mathbf{f}_p^{*\mathsf{T}} \boldsymbol{\beta}_j + \boldsymbol{\epsilon}_j $$
with $\boldsymbol{\beta}_j$ being the parameter vector of the model for wavelength $j$, $\beta_{0,j}$ the intercept term, and $\boldsymbol{\epsilon}_j$ the error term.

The parameters $\beta_{0,j}$ and $\boldsymbol{\beta}_j$ of the regression model were determined using least-squares estimation, which requires the estimation of the covariance matrix of $\mathbf{f}_p^*$. Because of the relatively large size of $\mathbf{f}_p^*$, we estimated the covariance matrix with the shrinkage method from \cite{LEDOIT2004365}, which computes the shrinkage coefficient explicitly. The parameters were found using all examples from the training set. Following this, we used the regression models to predict all relative radii of the planets $p$ in the test set with wavelength $j$:
$$\hat{y}_{p,j} = \beta_{0,j} + \mathbf{f}_p^{*\mathsf{T}} \boldsymbol{\beta}_j$$
The predicted relative radii $\hat{y}_{p,j}$ were re-indexed to the original indices $\hat{y}_{i,j}$ by copying $\hat{y}_{p,j}$ to all corresponding stellar spot noise instances and photon noise instances.

The only hyperparameters in our model were the percentiles used for estimating the minima of the light curve dips. We found these parameters by trial and error and refrained from fine tuning them further.

\subsection{TU Dortmund University} 

The team from TU Dortmund University, consisting of researchers working on applying machine learning algorithms in astro-particle physics, landed the 5th place on the leaderboard, going under the alias `Basel321' during the Challenge.
Their implementation is publicly available\footnote{Solution by \emph{TU Dortmund University} (Ranked 5th): \url{https://bitbucket.org/zagazao/ecml-discovery-challenge}}. It embraces three central ideas:
i) the preprocessing simplifies the input time series, yet retains much of their information in auxiliary features;
ii) the baseline architecture is largely retained, but consists of 2 input branches: one using as input these auxilliary features and the other using as inputs the stellar and planetary parameters; and
iii) a bagging ensemble is created, in which each member is trained on data that have undergone slightly altered preprocessing.

\begin{figure}
  \centering
  \includegraphics[scale=1]{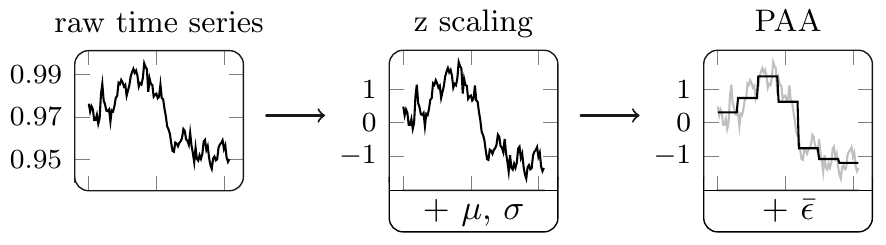}
  \caption{The TU Dortmund University team (Ranked 5th) simplifies the raw data with piecewise aggregate approximations (PAA) of z-scaled time series. The information lost during these transformations is retained in auxiliary features. Namely, the z-scaling produces time series with zero mean and unit variance, but the original means $\mu\in\mathbb{R}$ and variances $\sigma\in\mathbb{R}$ of each channel and observation are kept. The PAA consists of only one average value in each equi-sized segment, but the overall reconstruction errors $\bar{\epsilon}\in\mathbb{R}$ are maintained.}
  \label{fig:dortmund_paa}
\end{figure}

\begin{figure}
  \centering
  \includegraphics[scale=1]{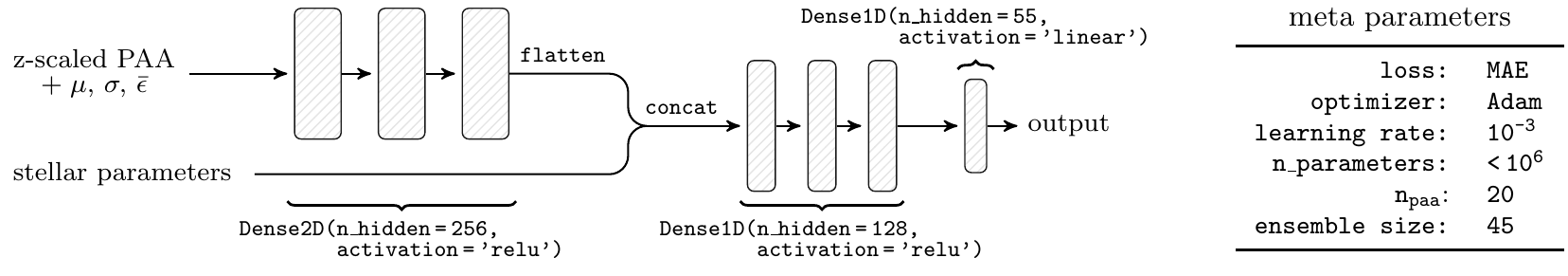}
  \caption{The first three layers of model used by the TU Dortmund University team (Ranked 5th) derive abstract features from each time series that is represented by a PAA. The auxiliary features $\mu$, $\sigma$ and $\bar{\epsilon}$ and the stellar parameters are also fed into the network. The last four layers combine these different kinds of inputs. A randomized parameter search has been employed to tune the number of layers and their size.}
  \label{fig:dortmund_architecture}
\end{figure}

\subsubsection{Preprocessing}
Figure~\ref{fig:dortmund_paa} shows how the input data are simplified by the use of z-scaled piecewise aggregate approximations (PAA) \cite{keogh_pazzani:2000}, of which the lost information is retained in the auxiliary features $\mu$, $\sigma$ and $\bar{\epsilon}$. These features describe each time series on a global level, while the PAA output captures the local shape. Namely, the PAA output is simply the average flux value in each of $n_{\text{paa}}$ equal-sized segments. The z-scaled PAA representation facilitates learning due to the decreased number of dimensions and due to the uniform scale in each dimension. These properties are particularly relevant in dense neural networks like the baseline solution, which can suffer from a large number of model parameters if the input dimension is large.

\subsubsection{Model architecture and training}
A fully connected DNN is trained on the extracted features and the planetary and stellar parameters. The architecture used, shown in Figure~\ref{fig:dortmund_architecture}, is similar to the baseline, but it includes one branch for the auxiliary features and one for the planetary and stellar parameters. Figure~\ref{fig:dortmund_architecture} also lists the associated hyperparameters.

Multiple instances of the above architecture, were then combined in a bagging ensemble. To increase the diversity, each ensemble member shifted its input by a different number $n \in [0, n_{\text{paa}})$ of time steps. This alteration is performed already before the preprocessing, so that each ensemble member uses different PAA segments. The final prediction was the median among all ensemble members' predictions.

\subsubsection{Observations}
Regarding the feature representation extracted in the preprocessing step, we observed the following:
i) a linear regression on the z-scaled PAA representation is already able to outperform the baseline solution;
ii) it is critical to maintain the information lost during this type of preprocessing -- this is achieved by the auxiliary features; and
iii) the use of shifting segments has remedied the fact that one set of PAA segments may not be optimal for all observations.

\section{What the winning models teach us}


We should stress again that the final score differences among the top-5 ranked solutions, as shown on Table \ref{table:leaderboard}, are statistically negligible and should thus be regarded as equivalent in terms of predictive power in our simulated data. Having clarified this, these solutions provide us with some interesting insights with regards to the problem.

 First of all, we observe that all 5 solutions make use of the additional stellar and planetary parameters (orbital period, stellar temperature, stellar surface gravity, stellar radius, stellar mass \& stellar $K$ magnitude). This shows that these features indeed contain relevant information for uncovering the transit depths in light curves contaminated by the presence of stellar spots. Moreover, this information is not redundant given the noisy light curves.
 
 Another interesting observation is that most solutions involve the use of highly non-linear nonparametric or overparameterized\footnote{The term `nonparametric' applies to models that are not restricted to a predetermined number of parameters. They can therefore adjust their complexity to the data at hand. Ensemble models can fall in this class. The term `overparameterized' refers to parametric models having a number of learnable parameters that exceeds the number of datapoints. DNNs can fall in this class. Through appropriate use of regularization methods it is possible to avoid overfitting even when fitting models of such high complexity.} models w.r.t. the original features, like DNNs and/or ensembles of learners. More specifically, 4 out of 5 teams use deep learning approaches (\emph{SpaceMeerkat}, \emph{Major Tom}, \emph{BV Labs} \& \emph{TU Dortmund University}) and 3 out of 5 (\emph{Major Tom}, \emph{BV Labs} \& \emph{TU Dortmund University} teams) use ensemble learning methods. The \emph{Major Tom} team does not apply any preprocessing of the data provided beyond feature normalization, leaving all feature extraction to be implicitly performed by the DNN, using appropriate regularization techniques (batch normalization \& dropout) to prevent overfitting. 

In contrast to this, the \emph{IWF-KNOW} team relied on the extraction of non-linear features from the original inputs informed by domain knowledge. They then trained simple linear models in this new feature space. 

The above are indicative of the non-linear nature of the problem. They also showcase the flexibility of machine learning and computational statistics methods in building models that capture this nonlinearity. One can extract informative features given domain knowledge to capture it and then use simple and explainable models like linear regression trained on them. Alternatively, one can simply use powerful overparameterized models, like DNNs and ensemble methods to implicitly learn  transformations of the original feature space that are useful for the purposes of predicting the transit depth.

Extracting a small number of meaningful features informed by domain knowledge (\emph{IWF-KNOW} \& \emph{BVLabs}) or appropriately summarizing the light curve information using signal processing techniques (\emph{SpaceMeerkat} \& \emph{TU Dortmund University}) allows for simpler models to be trained in the lower-dimensional extracted feature space. This allows for faster training and can also ultimately reduce overfitting. 

A more detailed look into how the 5 solutions control for overfitting also reveals they follow quite different approaches. \emph{SpaceMeerkat} uses a CNN rather than a fully connected DNN to reduce the number of effective learnable parameters. \emph{Major Tom} uses a fully connected DNN but controls for its complexity via batch normalization, dropout and the use of an ensemble of trained DNNs, rather than a single model. \emph{BV Labs} also make extensive use of ensembling and their neural network learner also uses  batch normalization and dropout. The fact that they operate on a much lower dimensional feature space (only 171 features per datapoint) also aids in reducing overfitting. \emph{IWF-KNOW} use linear regression models, which are characterized with high bias (i.e. more prone to underfitting than overfitting). They also operate on a lower-dimensional space (1656 features per datapoint) and apply shrinkage. Last but not least, \emph{TU Dortmund University} makes use of an ensemble which is interestingly built on data having undergone slightly different preprocessing. Training on perturbed inputs results in making them more robust to overfitting.

Two of the top-5 teams (\emph{BV Labs} \& \emph{IWF-KNOW}) made use of the fact that the training data contained multiple
datapoints corresponding to the same planet under (10 different photon noise and 10 different stellar spot noise instances). They treated the two noise sources as independent and averaged these out or took the median to obtain less noisy light curves. This was a sensible thing to do and such a scenario would indeed occur if multiple observations of the same target were to be obtained.

Finally, ignoring outlier flux values via smoothing/downsampling the light curves (\emph{SpaceMeerkat}), clipping values above 1 (\emph{SpaceMeerkat} \& \emph{BVLabs}) or by extracting summary statistics from the light curve and using them as features (\emph{SpaceMeerkat}, \emph{BVLabs}, \emph{IWF-KNOW} \& \emph{TU Dortmund University}) proved a useful strategy in building more robust models. 


\section{Conclusions}


Correcting transit light curves for the effects of stellar spots is a challenging problem, progress in which can have a high impact on exoplanetary science and exoplanet atmosphere characterization in particular. 

The primary goal of the \emph{Ariel Mission’s 1st Machine Learning Challenge} was to investigate the existence of fully automated solutions to this task that predict the transit depth with a precision of the order of $10^{-5}$ with the use of machine learning and computational statistics methodologies. The secondary goal was to bridge the machine learning and exoplanetary science communities. As we saw, both of these goals were met with success.

The aim of this work is to serve as a starting point for further interaction between the two communities. We described the data generation, the problem outline and the organizational aspects of the Challenge. We intend this to serve as a reference for the organisation of future challenges in data analysis for exoplanetary science. In the interests of communicating the  modelling outcomes of the Challenge, we also presented, analyzed and compared the top-5 ranked solutions submitted by the participants.

As evidenced by the top-5 entries, the Challenge indeed attracted the interest of both exoplanetary scientists and machine learning experts. The participants cover an impressive breadth of academic backgrounds and the submitted solutions an equally impressive range of approaches, from linear regression to convolutional neural networks. 

The solutions obtained demonstrate that it is indeed feasible to fully automate the process of efficiently correcting light curves for the effect of stellar spots to the desired precision.  One key insight obtained is that additional stellar and planetary parameters (orbital period, stellar temperature, stellar surface gravity, stellar radius, stellar mass \& stellar $K$ magnitude) can greatly improve the derivation of correct transit depths in the face of stellar spots.

Good solutions can be obtained by a wide range of modelling methodologies. They include simple, easily interpretable models, like linear regression, built on features derived from clever feature engineering, informed by exoplanet science theory. Other solutions ammount to training complex machine learning models using deep learning or ensemble learning, which automate the extraction of useful features from minimally preprocessed --even raw-- data. In the latter case, especially for DNN models it is crucial to take measures to prevent overfitting. These can include dimensionality reduction, ensembling, use of convolutional filters, batch normalization, dropout, training using perturbed data and combinations thereof.

The next steps of this work include refinement of the proposed solutions to handle more realistic simulated data, possibly involving both stellar spots and faculae (areas of the host star characterized by increased temperature). Upon successful performance on these, the provided solutions can then be used in the analysis pipeline of Ariel data or adapted to other instruments.

\section*{Acknowledgements}
N. Nikolaou acknowledges the support of the NVIDIA Corporation through the NVIDIA GPU Grant program. J. Dawson has received funding by the Science and Technology Facilities Council (STFC) as part of the Cardiff, Swansea \& Bristol Centre for Doctoral Training. He gratefully acknowledges the support of the NVIDIA Corporation with the donation of a Titan Xp GPU used for this research and the help of Nikki Zabel for the completion of this work. T.~Stepi\v{s}nik and M.~Petkovi\'{c} thank Martin Breskvar, Dragi Kocev and Nikola Simidjievski for their contributions, and Barbara Kra\v{s}ovec for her help with the Slovenian national supercomputing network. R. L. Bailey thanks the Austrian Science Fund (FWF) for funding: P31659-N27. The team from TU Dortmund University has been funded by the Federal Ministry of Education and Research of Germany as part of the competence center for machine learning ML2R (01IS18038A), and by Deutsche Forschungsgemeinschaft (DFG) within the Collaborative Research Center SFB 876 ``Providing Information by Resource-Constrained Data Analysis", project C3. This project has received funding from the European Research Council (ERC) under the European Union's Horizon 2020 research and innovation programme (grant agreement No 758892, ExoAI).
Furthermore, we acknowledge funding by the Science and Technology Funding Council (STFC) grants: ST/K502406/1, ST/P000282/1, ST/P002153/1 and ST/S002634/1.

\bibliography{ARIEL_ML_Challenge}{}
\bibliographystyle{aasjournal}

\end{document}